\documentclass[useAMS,usenatbib]{mn2e}

\usepackage{times}
\PassOptionsToPackage{hyphens}{url}
\usepackage{natbib}
\usepackage{graphicx}
\usepackage{txfonts}
\usepackage{subfig}
\usepackage[countmax]{subfloat}
\usepackage{url}
\def\lesssim{\mathrel{\hbox{\rlap{\hbox{\lower4pt\hbox{$\sim$}}}\hbox{$<$}}}}
\def\gtrsim{\mathrel{\hbox{\rlap{\hbox{\lower4pt\hbox{$\sim$}}}\hbox{$>$}}}}
\newcommand{\mincir}{\raise
-2.truept\hbox{\rlap{\hbox{$\sim$}}\raise5.truept
\hbox{$<$}\ }}
\newcommand{\magcir}{\raise
-2.truept\hbox{\rlap{\hbox{$\sim$}}\raise5.truept
\hbox{$>$}\ }}
\newcommand{\siml}{\raise -2.truept\hbox{\rlap{\hbox{$\sim$}}\raise5.truept
\hbox{$<$}\ }}
\newcommand{\simg}{\raise -2.truept\hbox{\rlap{\hbox{$\sim$}}\raise5.truept
\hbox{$>$}\ }}
\newcommand{\be}{\begin{equation}}
\newcommand{\ee}{\end{equation}}
\newcommand{\ba}{\begin{eqnarray}}
\newcommand{\ea}{\end{eqnarray}}

\title[ShaSS optical catalogues]{Shapley Supercluster Survey: construction of the photometric catalogues and $i$-band data release.}
\author[A. Mercurio et al.]
{A. Mercurio$^{1}$\thanks{E-mail: mercurio@na.astro.it}, P. Merluzzi$^{1}$, G. Busarello$^{1}$,  A. Grado$^{1}$, L. Limatola$^{1}$, C. P. Haines$^{2}$, \and M. Brescia$^{1}$, S. Cavuoti$^{1,3}$, M. Dopita$^{4,5}$, M. Dall'Ora$^{1}$, M. Capaccioli$^{1,6}$,  N. Napolitano$^{1}$, \and K. A. Pimbblet$^{7,8}$\\
$^{1}$INAF-Osservatorio Astronomico di Capodimonte, Salita Moiariello 16 I-80131 Napoli, Italy\\
$^{2}$Departamento de Astronom{\'{\i}}a, Universidad de Chile, Casilla 36-D, Correo Central, Santiago, Chile\\
$^{3}$INAF-Osservatorio Astronomico di Trieste, Via Tiepolo 11 I-34143 Trieste, Italy\\
$^{4}$Research School of Astronomy and Astrophysics, Australian National University, Cotter Rd., Weston ACT 2611, Australia\\
$^{5}$Astronomy Department, Faculty of Science, King Abdulaziz University, PO Box 80203, Jeddah, Saudi Arabia\\
$^{6}$Dipartimento di Fisica, Università Federico II, Via Cintia I-80126 Napoli, Italy\\
$^{7}$Department of Physics and Mathematics, University of Hull, Cottingham Road, Kingston-upon-Hull, HU6 7RX, UK\\
$^{8}$School of Physics, Monash University, Clayton, Melbourne, Victoria 3800, Australia\\
}

\begin{document}

\date{Accepted 2015 August 17. Received 2015 August 14; in original form 2015 March 10}

\pagerange{\pageref{firstpage}--\pageref{lastpage}} \pubyear{}

\maketitle

\label{firstpage}

\begin{abstract}

The Shapley Supercluster Survey is a multi-wavelength survey covering
an area of $\sim$23 deg$^2$ ($\sim$ 260 Mpc$^2$ at z=0.048) around the
supercluster core, including nine Abell and two poor clusters, having
redshifts in the range 0.045-0.050. The survey aims to investigate the
role of the cluster-scale mass assembly on the evolution of galaxies,
mapping the effects of the environment from the cores of the clusters
to their outskirts and along the filaments. The optical ($ugri$)
imaging acquired with OmegaCAM on the VLT Survey Telescope is
essential to achieve the project goals providing accurate multi-band
photometry for the galaxy population down to m*+6. We describe the
methodology adopted to construct the optical catalogues and to
separate extended and point-like sources. The catalogues reach average
5$\sigma$ limiting magnitudes within a 3$\arcsec$ diameter aperture of
$ugri$=[24.4,24.6,24.1,23.3] and are 93\% complete down to
$ugri$=[23.8,23.8,23.5,22.0]\,mag, corresponding to $\sim$
m*$_r$+8.5. The data are highly uniform in terms of observing
conditions and all acquired with seeing less than 1.1\,arcsec
full width at half-maximum. The median seeing in $r$-band is 0.6\,arcsec, corresponding to
0.56\,kpc\,h$^{-1}_{70}$ at z=0.048. While the observations in the
$u$, $g$ and $r$ bands are still ongoing, the $i$-band observations
have been completed, and we present the $i$-band catalogue over the
whole survey area. The latter is released and it will be regularly
updated, through the use of the Virtual Observatory tools. This
includes 734,319 sources down to $i$=22.0\,mag and it is the first
optical homogeneous catalogue at such a depth, covering the central
region of the Shapley supercluster.

\end{abstract}


\section{Introduction}
\label{intro}

The main aim of the Shapley Supercluster Survey (ShaSS) is to quantify
the influence of hierarchical mass assembly on galaxy evolution and to
follow such evolution from filaments to cluster cores, identifying the
primary location and mechanisms for the transformation of spirals into
S0s and dEs.  The most massive structures in the local Universe are
superclusters, which are still collapsing with galaxy clusters and
groups frequently interacting and merging, and where a significant
number of galaxies are encountering dense environments for the first
time. The Shapley supercluster (hereafter SSC) was chosen because of
i) the peculiar cluster, galaxy and baryon overdensities
\citep{sca89,ray89,fab91,ray91,def05}; ii) the relative dynamical
immaturity of this supercluster and the possible presence of infalling
dark matter haloes as well as evidence of cluster-cluster mergers
\citep[e.g.][]{bar94,qui95,bar98a,bar98b,kul99,qui00,bar01,dri04}; iii) the
possibility that it is the most massive bound structure known in the
Universe, at least in the 10\,Mpc central region
\citep[see][]{pea13}. These characteristics make the SSC an ideal
laboratory for studying the impact of hierarchical cluster assembly on
galaxy evolution and to sample different environments (groups,
filaments, clusters). Furthermore, the redshift range of this
structure \citep[0.033$<z<$0.060][]{qui95,qui97,pro06} makes it feasible to
measure the properties of member galaxies down to the dwarf regime,
providing that the observations reach the suitable depth. A detailed
discussion of the scientific aspects of the survey is given in
Merluzzi et al. (\citeyear{mer15}).

Although, the SSC has been investigated by numerous authors since its
discovery (Shapley \citeyear{sha30}) both for its cosmological
implications \citep[][and references
  therein]{sca89,ray89,pli91,qui95,koc04,fei13} and for studies of
cluster-cluster interactions
\citep[e.g.][]{kul99,bar00,fin04,ros05,mun08}, none of them could
systematically tackle the issue of galaxy evolution in the
supercluster environment due to the lack of accurate and homogeneous
multi-band imaging covering such an extended structure. ShaSS aims to
fill this gap measuring the integrated [magnitudes, colours, star formation rates (SFRs)] and
internal (morphological features, internal colour gradients)
properties of the supercluster galaxies.

ShaSS will map a region of $\sim$ 260 h$_{70}^{-2}$ Mpc$^2$ (at
z=0.048), centred on the SSC core, which is constituted by three Abell
clusters: A\,3558 (z=0.048, Melnick \& Quintana \citeyear{mel81},
Metcalfe, Godwin \& Spenser \citeyear{met87}; Abell richness $R$=4,
Abell, Corwin \& Olowin \citeyear{abe89}), A\,3562 (z=0.049, $R$=2) and
A\,3556 (z=0.0479, $R$=0); and two poor clusters SC\,1327-312 and SC
1329-313. The present survey covers also six other Abell
clusters: A\,3552, A\,3554, A\,3559, A\,3560, AS\,0724, AS\,0726, as
shown in Fig.~\ref{fig1}. The survey boundaries are chosen not only to cover
all 11 clusters and the likely connecting filaments, but also to
extend into the field and to map the structures directly connected to
the SSC core. In \citet{mer15} we derived the stellar mass density
distribution based on supercluster members showing that all the
clusters in the ShaSS area are embedded in a common network and
identified a filament connecting the SSC core and the cluster A\,3559
as well as the less pronounced overdensity extending from the SSC core
towards A\,3560.

The data set of the survey includes optical ($ugri$) and NIR ($K$)
imaging acquired with (VST) and Visible and Infrared Survey Telescope for Astronomy (VISTA) respectively, and optical
spectroscopy with AAOmega. At present the $i$-band imaging and AAOmega
spectroscopic surveys are completed, while the other observations are
ongoing. In addition, the recent public release of data from the
Wide-field Infrared Survey Explorer \citep[$WISE$,][]{wri10} provides
photometry at both near-IR (3.4,4.6\,$\mu$m) and mid-IR
(12,22\,$\mu$m) wavelengths, allowing independent measurements of
stellar masses down to $\mathcal{M}$=10$^{9}$M$_{\odot}$ at
10$\sigma$ and SFR down to
0.46\,M$_\odot$yr$^{-1}$ at 10$\sigma$ (0.2\,M$_\odot$yr$^{-1}$ at
5$\sigma$). Finally, in the 2--3\,deg$^2$ of the SSC core, panoramic imaging
in the UV (Galaxy Evolution Explorer, GALEX), optical (ESO Wide Field
Imager, WFI), NIR (UKIRT/WFCAM) and mid-infrared (Spitzer/MIPS) are
also available \citep{mer06,mer10,hai11}.

The optical survey, whose coverage is indicated by the 1\,deg$^2$
boxes in Fig.~\ref{fig1}, will enable us to i) derive accurate
morphologies, structural parameters ($\delta${\it log}$r_e\sim0.04$
and $\delta n_{Ser} \sim1$) as well as detect some of the observational
signatures related to the different processes experienced by
supercluster galaxies (e.g. extraplanar material); ii) estimate
accurate colours, photo-zs \citep[$\delta z<0.03$, see][]{chr12} and
stellar masses; iii) evaluate the SFRs and resolve the star forming
regions at least for the subsample of brighter galaxies. 

The survey depth enables global and internal physical properties
of Shapley galaxies to be derived down to $m^\star$+6.  In
  the first case of obtaining accurate measurements of aperture
  photometry and colours, we require signal-to-noise ratios (SNR) of
  20 in all four bands for SSC galaxies down to
  $m^\star$+6. Secondly, for the morphological analysis and resolving
  internal properties and structures there is a more stringent
  requirement of SNR$\sim$100 \citep[in a 3$\arcsec$ diameter
    aperture, see][]{con00,hau07} for the deeper $r$-band imaging. For
  this reason we are collecting the $r$-band imaging under the best
  observing conditions, with a full width at half-maximum (FWHM)$\sim$0.8\,arcsec or better,
  corresponding to 0.75\,kpc at z=0.048. Additionally, the $r$ imaging
  is fundamental to our weak lensing analysis, to ensure a sufficient
  density of lensed background galaxies with shape measurements.

With these data it will be possible to separate the different
morphological types, trace ongoing SF, reveal recent interaction or
merging activities and thus obtain a census of galaxies whose
structure appears disturbed by the environment \citep[e.g.][and
references therein]{sca07,mun09,kle14,hol14,lot08,lot11}. 

In order to achieve the scientific goals of the survey, accurate
  photometry is required. This implies a {\it clean} source catalogue
  containing, together with the measured photometric properties,
  indicators of the reliability of these measurements. This paper
  describes the methodology used to produce the photometric catalogues
  and the adopted procedures.

Observations are overviewed in Sect.~\ref{sec:2}. In Sect.~\ref{sec:3}
we describe the construction of the catalogues, the criteria to
classify spurious objects and unreliable detections, the flags
adopted in the catalogues and the procedure for star/galaxy
separation. The accuracy and completeness of the derived photometry is
discussed in Sect.~\ref{sec:4}. Each parameter of the released $i$-band
catalogues is detailed in Sect.~\ref{sec:5} and the summary is given in
Sect.~\ref{sec:6}.

Throughout the paper, we assume a cosmology with
$\Omega_{\mathrm{m}}$= 0.3, $\Omega_{\mathrm{\Lambda}}$= 0.7 and
H$_0$= 70 km s$^{-1}$ Mpc$^{−1}$. According to this cosmological model
1\, arcmin corresponds to 56.46 kpc at z=0.048. The magnitudes are given in
the AB photometric system.

\section{VST observations and data reduction}
\label{sec:2}

The optical survey (PI: P. Merluzzi) is being carried out using the
Italian INAF Guaranteed Time of Observations (GTO) with OmegaCAM on
the 2.6m ESO VLT Survey Telescope located at Cerro Paranal
(Chile). The Camera has a corrected field of view of 1$^{\circ}\times$1$^{\circ}$,
corresponding to $\sim$ 3.4$\times$3.4 h$_{70}^{-2}$ Mpc$^{2}$ at the
supercluster redshift, sampled at 0.21\,arcsec per pixel with a
16k$\times$16k detector mosaic of 32 CCDs, with gaps of 25-85\,arcsec
in between chips \footnote{More details on the camera are available at \url{http://www.eso.org/sci/facilities/paranal/instruments/omegacam/inst.html}\ .}.

\begin{figure}
\centering
\includegraphics[scale=0.4]{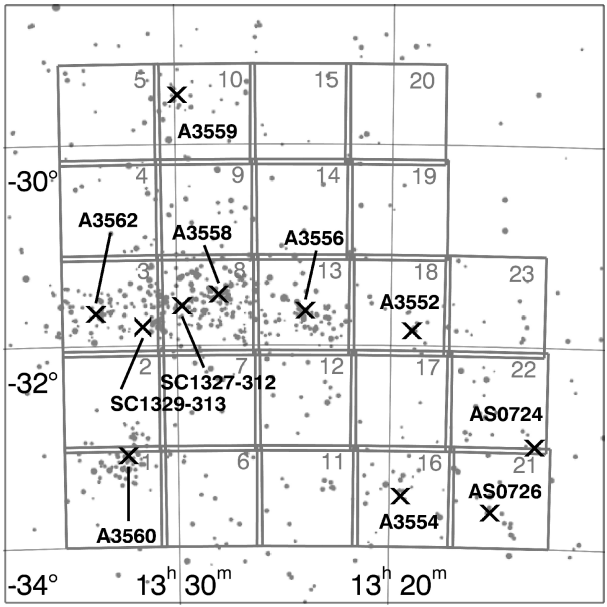}
\caption{VST fields mapping the ShaSS region. Dots indicate the
  supercluster members in the range v$_{heliocentric}$=13500-16000 km
  s$^{−1}$, taken from literature. Note that the literature redshift coverage is not uniform across the region shown. Black crosses show the cluster
  centres.}
\label{fig1}
\end{figure}

Each field is observed in four bands: $ugri$. To achieve the required
depth, total exposure times for each pointing are 2955s in $u$,
1400s in $g$, 2664s in $r$ and 1000s in $i$.  To bridge the gaps a
diagonal dither pattern of five exposures in $u$, $g$ and $i$, and
nine exposures in $r$ is performed, with step size of 25$\arcsec$
(15\,$\arcsec$ for the $r$ band) in X, and 85$\arcsec$ (45$\arcsec$ in
  $r$) in the Y-direction. The total area is covered by 23 contiguous
VST pointings overlapping by $\sim$3$\arcmin$ as shown in
Fig.~\ref{fig1}, where dots denote the spectroscopic supercluster
members ($13500 < V_h < 16000$\,km\,s$^{-1}$) available from
literature at the time of the survey planning. The X-ray centres are
indicated by crosses for all the known clusters except AS\,0726, whose
centre is derived by a dynamical analysis.

The survey started in 2012 February and will be completed in 2015
(spanning ESO periods P88-P95), provided that all the foreseen observations are carried out. At present, the survey coverage differs for each
band with only the $i$-band observations available for the whole
area. This implies that the results concerning the quality of the
photometry (depth, completeness, accuracy) are based on a
representative subsample of the final catalogues for $ugr$ (48\%, 43\% and
61\%, respectively) bands and for the whole catalogue in $i$ band.

\begin{figure}
\centering
\includegraphics[scale=0.4]{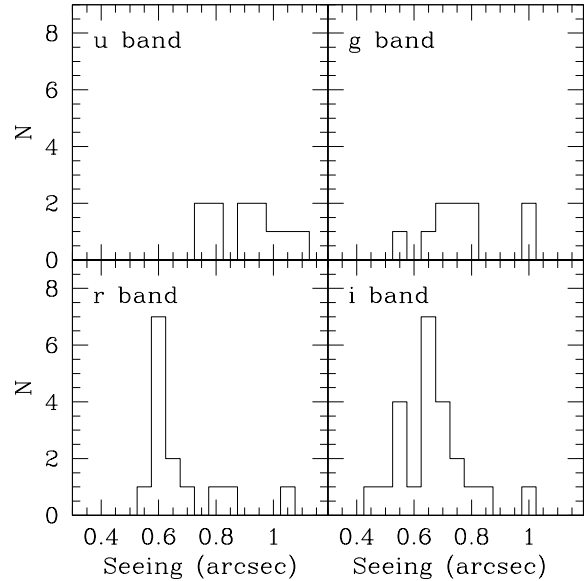}
\caption{Seeing distribution in the four observed bands (see text).}
\label{fig2}
\end{figure}

The data are collected on clear and photometric nights with good and
uniform seeing conditions. In Fig.~\ref{fig2} we plot the seeing
values of 11, 10, 14 and 23 fields in $ugri$, respectively. Out of the
observed fields about 80\% ($gi$) and 90\% ($r$) are acquired with
FWHM$\le$ 0.8\,arcsec, with the median seeing in $r$ band equal to
0.6\,arcsec, corresponding to 0.56 kpc h$^{-1}_{70}$ at z=0.048. The
$u$ band is characterized by a slightly poorer seeing. We discuss the
effect of the seeing on the aperture magnitudes in
Sect.~\ref{sec:3.1.2}.

The data reduction is already described in Merluzzi et
al. \citeyear{mer15}. Here we summarize the main steps
for the reader's convenience.

Images are reduced and combined using the VST--Tube imaging pipeline
(Grado et al. \citeyear{gra12}), developed {\it ad hoc} for the VST
data. The pipeline follows the standard procedures for bias
subtraction and flat-field correction. A normalized combination of the
dome and twilight flats, in which the twilight flat is passed through
a low-pass filter first, were used to create the master flat.  A gain
harmonization procedure has been applied, finding the relative CCD
gain coefficients which minimizes the background level differences in
adjacent CCDs. A further correction is applied to account for the light
  scattered by the telescope and instrumental baffling. This is an
  additional component to the background, which, if not corrected for,
  causes a position-dependent bias in the photometric measurements. This
component is subtracted through the determination and the
application of the illumination correction (IC) map. The IC map is
determined by comparing the magnitudes of photometric standard fields
with the corresponding SDSS-DR8 (Sloan Digital Sky Survey-Data Release
8) PSF magnitudes.

For the $i$ band a correction is required because of the fringe pattern
due to thin-film interference effects in the detector from sky
emission lines.  The fringing pattern is estimated as the ratio between
the Super-Flat and the twilight sky flat, where Super-Flat is obtained
by overscan and bias correcting a sigma clipped combination of science
images. The fringe pattern is subtracted from the image, applying a
scale factor which minimizes the absolute difference between the peak
and valley values (maximum and minimum in the image background) in the
fringe corrected image.

The photometric calibration onto the SDSS photometric
system is performed in two steps: first a relative photometric
calibration among the exposures contributing to the final mosaic image
is obtained through the comparison of the magnitudes of the bright
unsaturated stars in the different exposures, using the software SCAMP
(Bertin \citeyear{ber06})\footnote{Available at
  \url{http://www.astromatic.net/software/scamp}\ .}; then the absolute
photometric calibration is computed on the photometric nights
comparing the observed magnitude of stars in photometric standard
fields with SDSS photometry. For those fields observed
on clear nights, we take advantage of the sample of
bright unsaturated stars in the overlapping region between clear and
photometric pointings and by using SCAMP, each exposure of the clear
fields is calibrated on to the contiguous photometrically calibrated
field. The magnitude is then calibrated by adopting the following
relation:

\begin{equation}
    M' = M + \gamma C + AX + ZP,
\end{equation}
where $M$ is the magnitude of the star in the standard system, $M'$ is
the instrumental magnitude, $\gamma$ is the coefficient of the colour
term, $C$ is the colour of the star in the standard system, $A$ is the
extinction coefficient, $X$ is the airmass and $ZP$ is the zero-point. The results are reported in Table~\ref{tab1}.

\begin{table}
\centering
\small
\begin{tabular}{ l l l l l }
\hline
\hline
Band & Colour term & Colour & Extinction & Zero Point\\
     &             &        & Coefficient&           \\ 
\hline
  $M$  & $\gamma$    & $C$      & $A$          & $ZP$\\
\hline
$u$ & 0.026$\pm$0.019 & u-g & 0.538 &  23.261$\pm$0.028 \\
$g$ & 0.024$\pm$0.006 & g-i & 0.180 &  24.843$\pm$0.006 \\
$r$ & 0.045$\pm$0.019 & r-i & 0.100 &  24.608$\pm$0.007 \\
$i$ & 0.003$\pm$0.008 & g-i & 0.043 &  24.089$\pm$0.010 \\
\hline
\hline
\end{tabular}
\caption{Absolute photometric calibration coefficients.}
\label{tab1}
\end{table}

\begin{figure}
\centering
\includegraphics[scale=0.3]{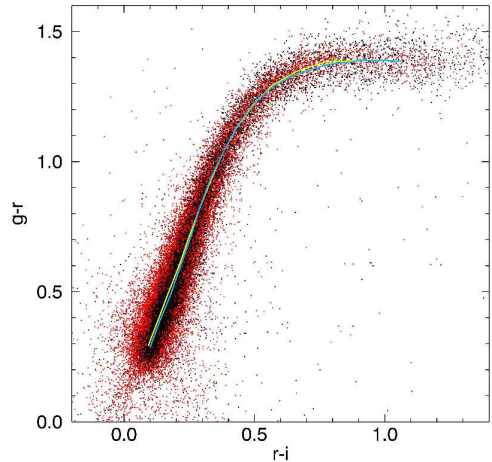}
\caption{Colour-colour ($g$-$r$,$r$-$i$) diagram showing a sample of stars
  from ShaSS (black dots) overimposed on a sample of SDSS stars (red
  dots). The curves trace the loci of the stars in the plane (ShaSS:
  cyan; SDSS: yellow). The distance between the curves never exceeds
  $\sim$0.015\,mag.}
\label{fig3}
\end{figure}

To further check our photometric calibration, in particular for stars
in fields observed in clear nights, we derive the median stellar loci
of SDSS and ShaSS stars in ($g-r$, $r-i$) colour-colour space (see
Fig.~\ref{fig3}). The distance between the two loci, over the whole
range of colours, never exceeds 0.015 mag, which is less than the
expected error on colours.

When photometric calibration is performed, the background is removed with
the software SWARP (Bertin et al. \citeyear{ber02})\footnote{Available
  at \url{http://www.astromatic.net/software/swarp}\ .}, which also
standardizes the zero-point to a value of 30.0 mag.

\section{Catalogue construction and star/galaxy classification}
\label{sec:3}

The procedure adopted for the source extraction is optimized for the goals
of the survey. Due to the image depth, sources span a wide range of size,
luminosity and morphology, and thus we need a multifaceted approach to obtain
robust measures of their aperture and total magnitudes. Moreover, the
SSC is located at relatively low galactic latitude
($b\sim$30), which implies the presence of a large number of stars across the
survey area, making the star/galaxy classification a crucial issue for
the catalogue's construction.

The photometric catalogues are produced using the software SExtractor
\citep{bertin96} in conjunction with PSFEx\footnote{Available at
  \url{http://www.astromatic.net/software/psfex}\ .}
(\citealp{bertin11}), which performs PSF fitting photometry. We
extract independent catalogues in each band, which are then matched
across the four wavebands using STILTS (Taylor \citeyear{tay06}).

\subsection{Catalogue construction}
\label{sec:3.1}

The procedure adopted for source extraction aims: \textit{i)} to
detect as many sources as possible, while minimizing the contribution
from spurious objects, \textit{ii)} to produce accurate measurements
of positions and photometric quantities, \textit{iii)} to flag objects
in the haloes of bright stars and hence could have had their
  photometry affected. During the construction of the catalogues, the
results have been always visually inspected on the images to check the
residual presence of spurious objects or misclassified objects, like
traces of satellites, fake objects due to cross-talk, effects of bad
columns.

\subsubsection{Source detection}
\label{sec:3.1.1}

Sources included in the final catalogue are extracted in four steps:
\textit{i)} sky background modelling and subtracting, \textit{ii)}
image filtering, \textit{iii)} thresholding and image segmentation,
\textit{iv)} merging and/or splitting of detections.

In order to optimize the automatic background estimation, we
obtained catalogues by adopting different {\tt{BACK\_SIZE}} and
{\tt{BACK FILTERSIZE}} and compared sources extracted in each
catalogue, both in terms of number of spurious detections and
photometric quantities, such as aperture and Kron magnitudes and flux
radius. Finally, given the average size of the objects, in pixels, in
our images, and in order to minimize the number of spurious
detections, we set the {\tt{BACK\_SIZE}} and {\tt{BACK FILTERSIZE}}
to 256 and 4 respectively, for all fields and bands.  To get accurate
background values for the photometry, the background is also
recomputed in an area centred around the object in question, setting
{\tt{BACKPHOTO\_TYPE}} to LOCAL and the thickness of the background
LOCAL annulus ({\tt{BACKPHOTO\_THICK}}) to 24.

Once the sky background is subtracted, the image must be filtered to
detect sources. To determine whether the Gaussian or top-hat filter
was optimal for our scientific objectives, a specific analysis was
carried out on the $g$-, $r$- and $i$-band images for field 8. The
catalogues produced using the top-hat filter were found to contain
more spurious sources than those with the Gaussian filter, while the
photometric measurements as well as the completeness of the catalogues
were confirmed to be equivalent. Finally we chose to apply the
Gaussian filter ({\tt{gauss\_3.0\_5x5.conv}}).

The detection process is mostly controlled by the thresholding
parameters ({\tt{DETECT\_THRESHOLD}} and
{\tt{ANALYSIS\_THRESHOLD}}). The choice of the threshold must be
carefully considered. A too high threshold results in the loss of a
high number of sources in the extracted catalogue, while a too low
value leads to the detection of spurious objects. Hence, a compromise
is needed by setting these parameters according to the image
characteristics, the background rms, and also to the final scientific
goal of the analysis.

\begin{figure*}
\centering
\subfloat[]{
\includegraphics[scale=0.4, bb= 30 187 539 581,clip]{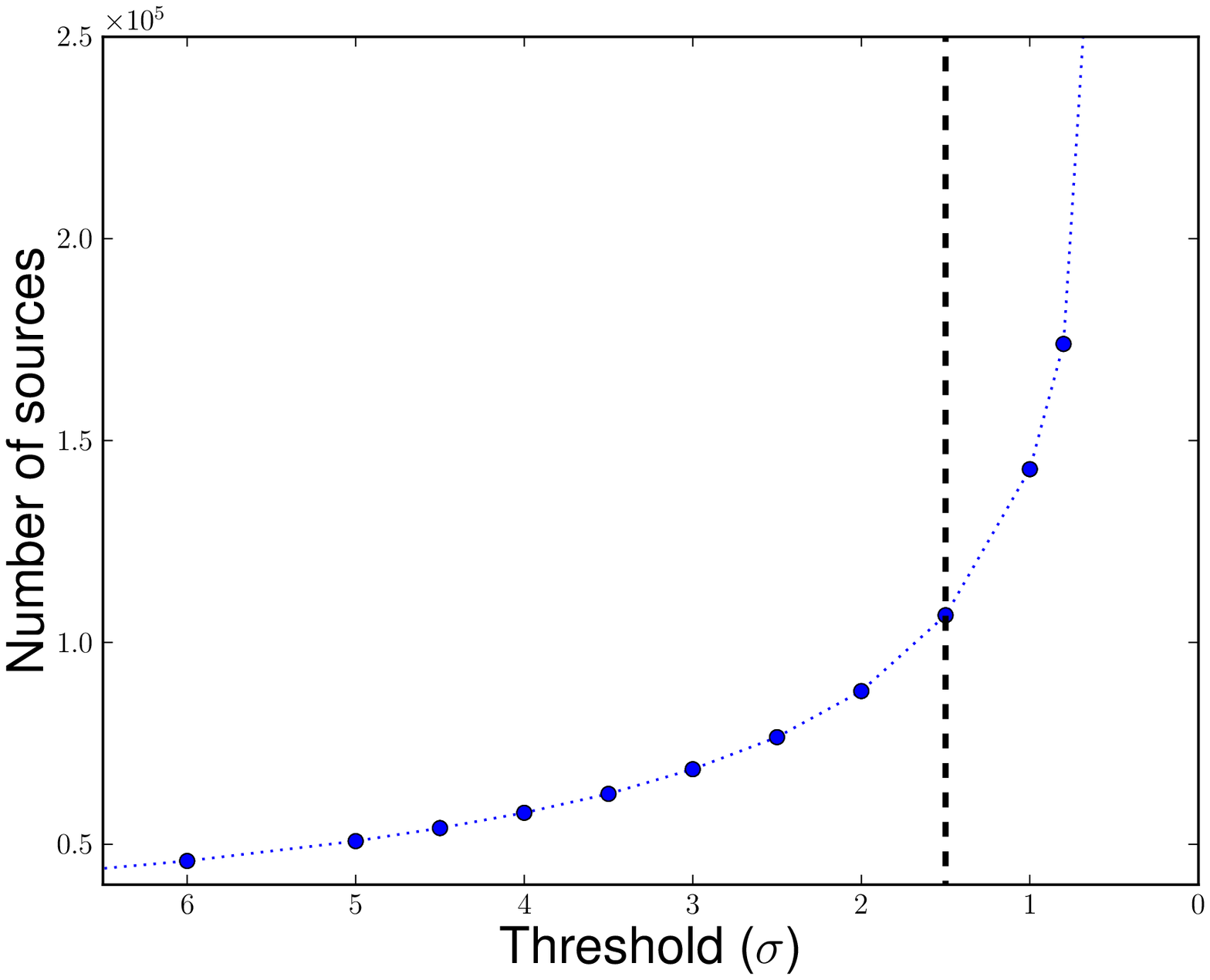}
}
\subfloat[]{
\includegraphics[scale=0.4, bb= 30 187 539 581,clip]{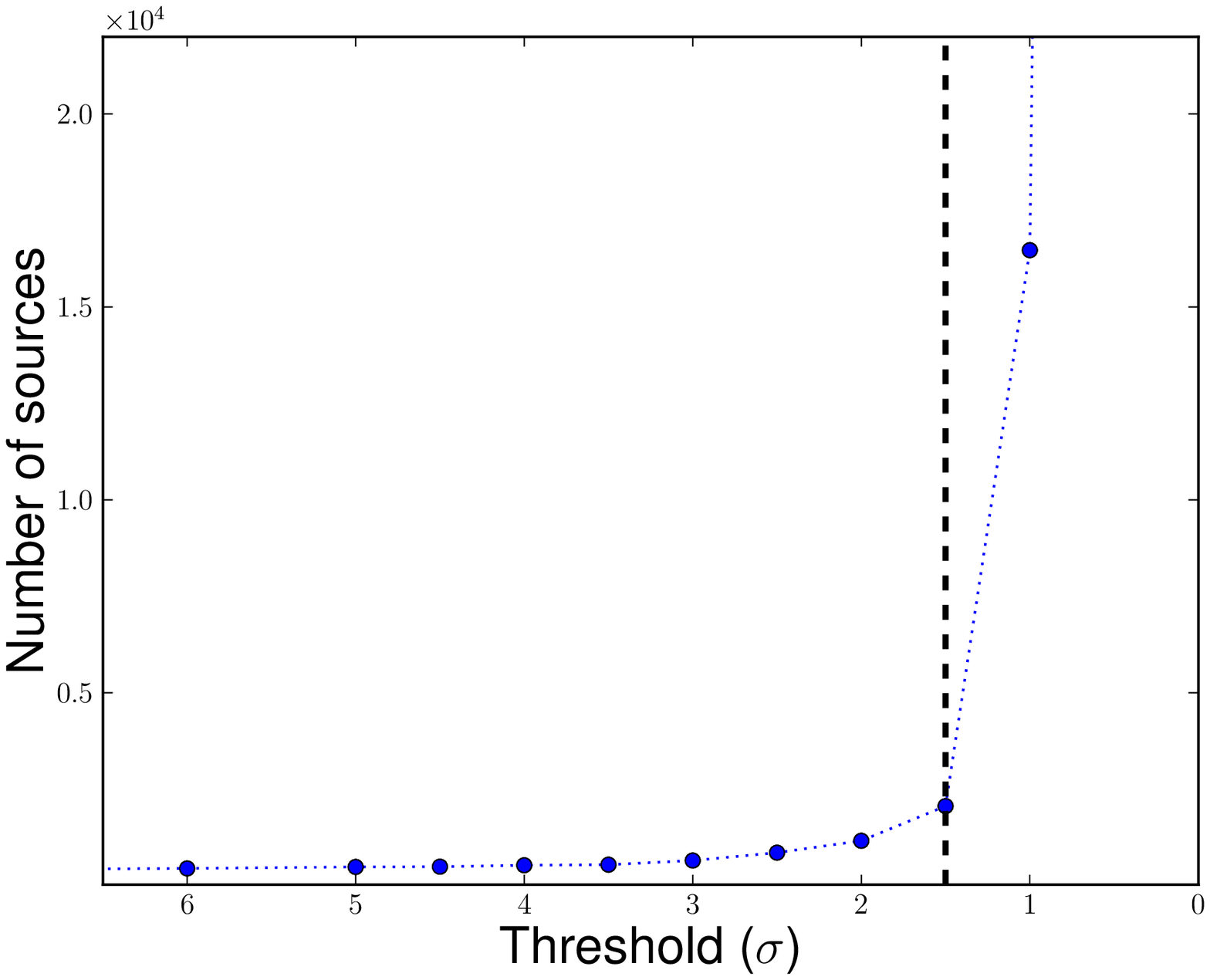}
}
\caption{Number of extracted sources for the $r$-band image ({\it left-hand panel}) and negative image ({\it right-hand panel}, see text) of field 12 as a function of the SExtractor threshold.
The dashed vertical line indicates in both panels the adopted detection threshold.
}
\label{fig4}
\end{figure*}

For our analysis, the threshold value was chosen to maximize the
number of detected sources, while simultaneously keeping the
  number of spurious detections to a minimum. To identify the optimal
  threshold value, we counted the number of extracted sources for
  different threshold values in 1\,deg$^2$ $r$-band image and the
  corresponding negative image, which is the scientific image
  multiplied by -1. As the threshold decreases, the number of detected
  sources (real plus spuriuos detections) increases, as shown in the
  left-hand panel of Fig.~\ref{fig4}.  In the negative image (right panel
  of Fig.~\ref{fig4}) the number of sources (spurious detections in
  this case) increases smoothly down to a certain value of the
  threshold, beyond which it shows a dramatic change in
  steepness. The suitable threshold value corresponds to this change
in the trend of source number counts. As pointed out above, the
catalogue was visually inspected to avoid residual spurious detections
and to verify the deblending parameters.

Two or more very  nearby sources can be mistakenly detected as a unique
connected region of pixels above threshold and, in order to correct
for this effect, SExtractor adopts a deblending method based on a
multi-thresholding process. Each extracted set of connected pixels is
re-thresholded at $N$ levels, linearly or exponentially spaced between
the initial extraction threshold and the peak value. Also here we
should find a compromise for the deblending parameter, since too
high a value leads to a lack of separation between close sources, while too low a value leads to split extended spiral galaxies into
several/multiple components.

In general, the choice of the deblending (but also the threshold)
parameter is related to the main scientific interest. If we are
mostly interested in the analysis of faint and small objects, we have
to fix low values for deblending and threshold parameters, at the cost
of splitting up the occasional big bright spiral galaxy into many
pieces. On the other hand, by fixing larger values for deblending and
threshold parameters, we correctly deblend the larger objects, but
we lose depth. However, the present data cover large areas and
the science involves the analysis of both large and bright as well as
faint and small galaxies.
 
\begin{table}
\centering
\small
\begin{tabular}{ l r }
\hline
\hline
\rule[-1.0ex]{0pt}{2.5ex}  Parameter & Values\\
\hline
{\tt{DETECT\_MINAREA}} & 5\\
{\tt{DETECT\_THRESH}} & 1.5$\sigma$ \\
{\tt{ANALYSIS\_THRESH}} & 1.5$\sigma$ \\
{\tt{DEBLEND\_NTHRESH}} & 32\\
{\tt{DEBLEND\_MINCONT}}$_{hot}$ & 0.001\\
{\tt{DEBLEND\_MINCONT}}$_{cold}$ & 0.01\\
{\tt{BACK\_SIZE}} & 256\\
{\tt{BACK\_FILTERSIZE}} & 4\\
\hline
\end{tabular}
\caption{Main input parameters set in the SExtractor configuration files. {\tt{DETECT\_MINAREA}} and {\tt{BACK\_SIZE}} are expressed in pixels.}
\label{tab2}
\end{table}

For this reason it is not possible to fix a unique value for threshold
and deblending parameters, and we used a two-step approach in the
catalogue extraction (e.g.: Rix et al. \citeyear{rix04}, Caldwell et
al. \citeyear{cal08}). First, we ran SExtractor in a so called {\it
  cold mode} where the brightest and extended sources are properly
deblended; then, in a second step, we set configuration parameters in
order to detect fainter objects and to properly split close sources
({\it hot mode}). We combined the two catalogues by replacing extended
objects, properly deblended in {\it cold mode}, in the catalogue of
sources detected in the {\it hot mode}, and by deleting multiple
detections of these extended sources. In order to combine these
  two catalogues we compare the segmentation maps obtained with $cold$
  and $hot$ modes. First we flagged the different objects detected in
  the hot mode, but lying inside the Kron area of the same object
  detected in cold mode, which are typically 80--120 per field. These
  objects are, in general, large spiral galaxies, galaxies with
  superimposed smaller galaxies or stars and close objects not
  properly deblended in {\it cold mode}. From the visual inspection of
  all the flagged objects we chose {\it cold mode} detection for
  spiral galaxies and {\it hot mode} for other cases.  The main
values set for SExtractor are listed in Table~\ref{tab2}.

\subsubsection{Aperture photometry and total magnitudes}
\label{sec:3.1.2}

Measurements of position, geometry, and photometric quantities are
included in the final catalogues for all the detected and properly
deblended sources. Among the photometric quantities calculated are:
aperture magnitudes ({\tt{MAG\_APER}}), measured in 10 circular
apertures whose fixed diameters are reported in Table~\ref{tab3};
isophotal magnitudes ({\tt{MAG\_ISO}}), computed by considering the
threshold value as the lowest isophote and using two different
elliptical apertures: the Kron ({\tt{MAG\_AUTO}}, Kron
\citeyear{kron80}) and the Petrosian ({\tt{MAG\_PETRO}}, Petrosian
\citeyear{petro76}) magnitudes, which are both estimated through an
adaptive elliptical aperture. The adopted Kron and Petrosian factors,
and the minimum radius fixed for the elliptical apertures are reported
in Table~\ref{tab3} ({\tt{PHOT\_AUTOPARAMS}} and
{\tt{PHOT\_PETROPARAMS}}, respectively).

In order to test the effect of seeing on the aperture magnitudes we
used the program SYNMAG (Bundy et al., \citeyear{bun12}). With this
software we can input magnitudes, effective radii and S{\'e}rsic indices
to obtain the synthetic magnitude of the source inside an aperture of
fixed radius.  We found that for photometry measured within fixed
apertures of 8\,$\arcsec$ diameter, the level of flux lost
remains below 0.02 mag for both exponential (n$_{Sersic}$=1)
and de Vaucouleurs (n$_{Sersic}$=4) profiles, for seeing levels in the
range 0.5-1.1\,arcsec and effective radii in the range
3.0-100.0\,arcsec.

\begin{table}
\centering
\small
\begin{tabular}{ l r }
\hline
\hline
\rule[-1.0ex]{0pt}{2.5ex}  Parameter & Values\\
\hline
{\tt{PHOT\_APERTURES}} & 1.5,3.0,4.0,8.0,16.0,30.0,45.0,90.0\\
{\tt{PHOT\_APERTURES}} & 3*FWHM,8*FWHM\\
{\tt{PHOT\_AUTOPARAMS}} & 2.5, 3.5\\
{\tt{PHOT\_PETROPARAMS}} & 2.0, 3.5\\
\hline
\end{tabular}
\caption{Adopted aperture diameters, {\tt{MAG\_AUTO}} and
{\tt{MAG\_PETRO}} parameters. {\tt{PHOT\_APERTURES}} are expressed
in arcsec (first row) and as a function of the measured FWHM (second
row).}
\label{tab3}
\end{table}

To describe the size of the sources we extract: the half flux radius,
i.e. the {\tt{FLUX\_RADIUS}} containing the 50\% of the total light
and that containing the 90\% of the light. We also measured the
{\tt{PETRO\_RADIUS}}, defined as the point in the radial light profile
at which the isophote at that radius is 20\% of the average surface
brightness within that radius, and the {\tt{KRON\_RADIUS}}, which is
the characteristic dimension of the ellipse used to calculate the
{\tt{KRON\_MAGNITUDE}}. The Kron and the Petrosian magnitudes are
measured within elliptical apertures, whose semi-major and semi-minor
axes are equal to {\tt{A\_IMAGE}} and {\tt{B\_IMAGE}} multiplied by
the {\tt{KRON\_RADIUS}} and {\tt{PETRO\_RADIUS}} parameters,
  respectively.

As position parameters, we measured the barycenter
coordinates both in pixels ({\tt{X\_IMAGE}}, {\tt{Y\_IMAGE}}) and in
degrees ({\tt{ALPHA\_J2000}}, {\tt{DELTA\_J2000}}), computed as the
first order moments of the intensity profile of the image, and the
position of the brightest pixel ({\tt{ALPHAPEAK\_J2000}},
{\tt{DELTAPEAK\_J2000}}).

\subsubsection{Model-derived magnitudes}
\label{sec:3.1.3}

By using the PSFEx tool \citet{bertin11} it is possible to model the
PSF of the images. We considered only non-saturated
point sources with a SNR 
 higher than 80. Spatial variations of the PSF were modelled with a
 third degree polynomial of the pixel coordinates. The
 main values set for PSFEx parameters are given in
 Table~\ref{tab4}.  More details on the PSF modelling with the PSFEx
 tool can be found \citet{bertin11},
 \citet{mohr12} and Bouy et al. (\citeyear{bou13}).

\begin{table}
\centering
\small
\begin{tabular}{ l r }
\hline
\hline
\rule[-1.0ex]{0pt}{2.5ex}  Parameter & Values\\
\hline

{\tt{BASIS\_TYPE}} & PIXEL\_AUTO\\
{\tt{BASIS\_NUMBER}} & 20\\
{\tt{PSF\_ACCURACY}} & 0.01\\
{\tt{PSF\_SIZE}} & 25,25\\
\hline
\end{tabular}
\caption{Main input parameters set in the PSFEx configuration file. {\tt{PSF\_SIZE}} are expressed in pixels.}
\label{tab4}
\end{table}

Hence, it is possible to run SExtractor taking the PSF models
as input, using them to carry out the PSF-corrected model fitting
photometry for all sources in the image. With a combination of
SExtractor and PSFEx, we obtained magnitudes from: {\it (i)} the PSF
fitting ({\tt{MAG\_PSF}} and the point source total magnitude
{\tt{MAG\_POINTSOURCE}}); {\it (ii)} the fit of a spheroidal component
({\tt{MAG\_SPHEROID}}); {\it (iii)} the fit of a disc component
({\tt{MAG\_DISK}}); and {\it (iv)} the sum of the bulge and the disc
components, centred on the same position, convolved with the local PSF
model ({\tt{MAG\_MODEL}}). We also extracted morphological parameters
of the galaxies, such as spheroid effective radius
({\tt{SPHEROID\_REFF\_IMAGE}}), spheroid S{\'e}rsic index
({\tt{SPHEROID\_SERSICN}}), spheroid aspect
({\tt{SPHEROID\_ASPECT\_IMAGE}}), disc scale length
({\tt{DISK\_SCALE\_IMAGE}}) and disc aspect
({\tt{DISK\_ASPECT\_IMAGE}}).

The model of the PSF is also helpful to get a more accurate
star/galaxy classification (see Sect.~\ref{sec:3.3}).

\subsection{Haloes and spikes of bright stars, and image borders}
\label{sec:3.2}

\begin{figure}
\centering
\includegraphics[scale=0.4]{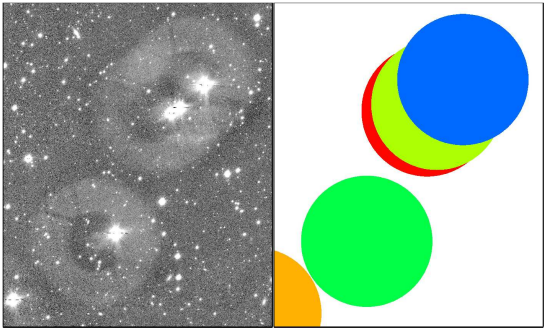}
\caption{ A crowding of five saturated stars with related haloes
  (left) and the corresponding halo mask (right). The $i$-band
  magnitudes of the stars are, left to right, 8.77, 8.24, 8.99, 8.56,
  7.81. A logarithmic stretch was applied to the image to increase the
  background. The image corresponds to an area of 9$\arcmin$ $\times$
  6$\arcmin$.}
\label{fig5}
\end{figure}

Multiple reflections in the internal optics of OmegaCAM can produce
complex image rings and ghosts (hereafter star haloes) near
bright stars \footnote{See
  \url{http://www.eso.org/observing/dfo/quality/OMEGACAM/qc/problems.html}\ .}.
These haloes are characterized by a central region in which the
surface brightness is depressed and an outer corona with enhanced
surface brightness, both characterized by an irregular pattern as
shown in Fig.~\ref{fig5} (left panel) for the $i$-band images. Those
haloes all have the same radius of 830 pixels in all bands and the
position of their centres is related to the position of
the parent star in the OmegaCAM field. The `intensity' (depth of the
inner depression and brightness of the corona) of the halo depends on
the band, being more manifest in the $i$ band and practically
negligible in $u$ band. Moreover, the intensity of the halo is related
to the magnitude of the parent star. To quantify this behaviour we
assigned a value proportional to the halo masks: mask=INT(100$\times$
i$_{mag}$). Notice that parent stars are saturated in our images, so
we draw their magnitudes from the USNO-B1.0 catalogue.
Whenever two or more haloes overlap, the mask assumes locally the
value corresponding to the halo of the brightest star.  We limited the
creation of masks to stars brighter than $i$=9.6 mag, although the
haloes produced by stars with $i >$ 9.0 mag already do not show
surface brightness excesses higher than $\sim$0.2 times the r.m.s. of
the background.  In all 23 VST fields in $i$ band, we identified and
masked 299 haloes.  In the right panel of Fig.~\ref{fig5} we show an
example of a halo mask.

\begin{figure}
\includegraphics[scale=0.5]{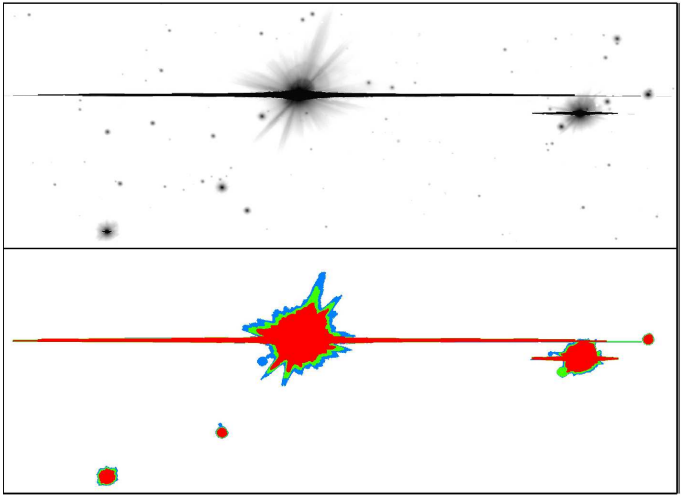}
\caption{An area containing a star with prominent spikes and other
  saturated stars (top) and the corresponding spike mask (bottom). The
  three levels of the mask correspond to three different surface
  brightness thresholds ($\mu_i <$ 21, 20.5, 20 mag arcsec$^{-2}$,
  coded here in blue, green and red respectively). The shown field
  measures 2.8$\arcmin$ $\times$ 7.7$\arcmin$.}
\label{fig6}
\end{figure}

The saturated stars with haloes also affect the images by
  producing spike features which are masked as follows. First we
identified the position of all saturated pixels according to the value
corresponding to the {\tt SATURATE} keyword in the header of each
image. Then we applied a region growth algorithm around each saturated
pixel to obtain a mask consisting of three levels, corresponding to
three threshold values of the surface brightness of 20, 20.5 and 21
mag arcsec$^{-2}$. An example of a spike mask is shown in
Fig.~\ref{fig6}.

The halo and spike masks have been used to flag each source s$_i$
considering the circular area embedding 50\% of the flux, A50$_i$. From
both halo and spike masks we derive a couple of flags: i) the fraction
of A50$_i$ affected by the halo and/or spike mask; ii) the halo and/or
mask value within the affected portion of A50$_i$.

\begin{figure}
\centering
\includegraphics[scale=0.5]{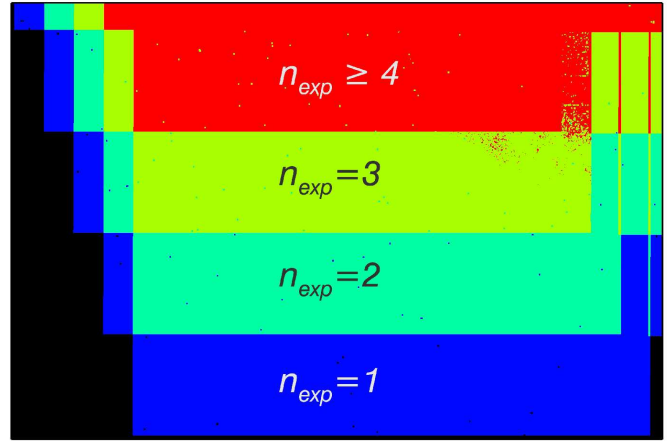}
\caption{Cutoff of an exposure mask near to a corner of a VST
  image. The area shown is about 14$\arcmin$ $\times$ 12$\arcmin$. The colours
  correspond to the different numbers of exposure as indicated.}
\label{fig7}
\end{figure}

The dither acquisition mode adopted for the observations produces a
non-uniform coverage of the co-added image, so that the catalogue
  of detections covering the whole mosaic image will contain
  detections from regions with different depths.  For a
five(nine)-dither pattern at least three(five) exposures cover almost the
whole image in $ugi$($r$). Along the borders, however, there are
stripes where only 1-2(1-4) exposures contribute to the final co-added
mosaic in $ugi$($r$). We decided to exclude these detections from the
final catalogues. This {\it cut} was possible thanks to the
3$^{\arcmin}$ overlaps among the contiguous 1 deg$^2$ VST fields,
which ensure that a common area (and common detections) is present
even after the borders are cut. To mask out we empirically identified
the external areas with different exposures by just associating a
range of values of the weight mask to a number of exposures. An
example of an exposure mask is given in Fig.~\ref{fig7}. Objects lying
in areas with mask values lower than 3(5) are not included in the
survey catalogues, except for a very few number of sources located in the gaps
among the detectors.

We also excluded those objects for which more than 80\% of their
pixels (within a circular region with radius equal to 50\% flux
radius) have zero weight.

\subsection{Star/Galaxy separation}
\label{sec:3.3}

To separate extended and point-like sources, we adopt a progressive
approach analogous to those described in Annunziatella et
al. (\citeyear{ann13}), using the following parameters provided by
SExtractor: (i) the \textit{stellarity index} ({\tt{CLASS\_STAR}}); (ii)
the half-light radius ({\tt{FLUX\_RADIUS}}); (iii) the new SExtractor
classifier {\tt{SPREAD\_MODEL}}; (iv) the peak of the surface
brightness above background (\texttt{$\mathrm{\mu_{max}}$}); (v) a
final visual inspection for objects classified as galaxies but with
borderline values of the stellarity index ({\tt{CLASS\_STAR}}$\ge$0.9; see below).

\begin{figure*}
\centering
\subfloat[]{
\includegraphics[scale=0.4, bb= 30 170 545 595,clip]{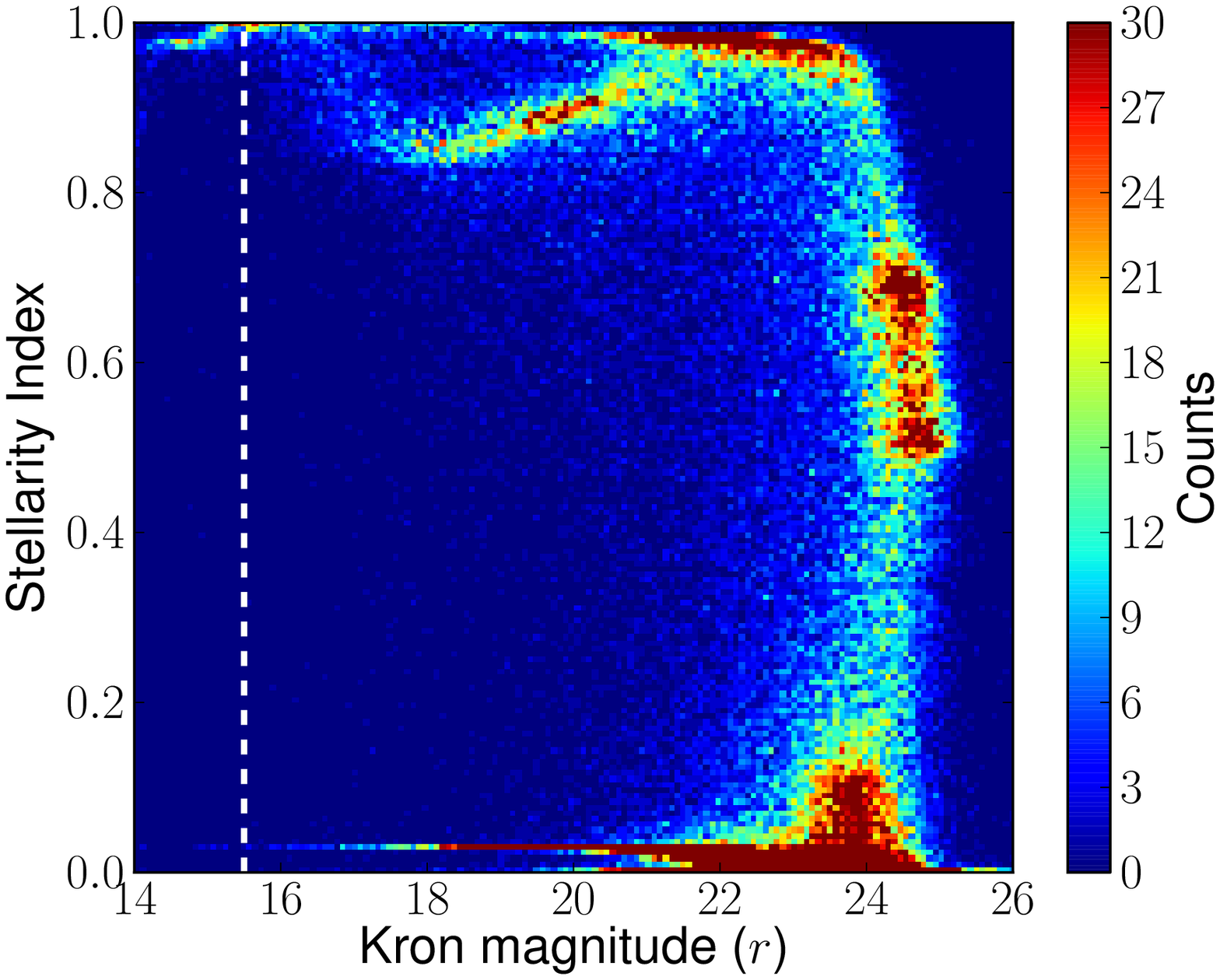}
\label{fig8a}}
\subfloat[]{
\includegraphics[scale=0.4, bb= 30 170 545 595,clip]{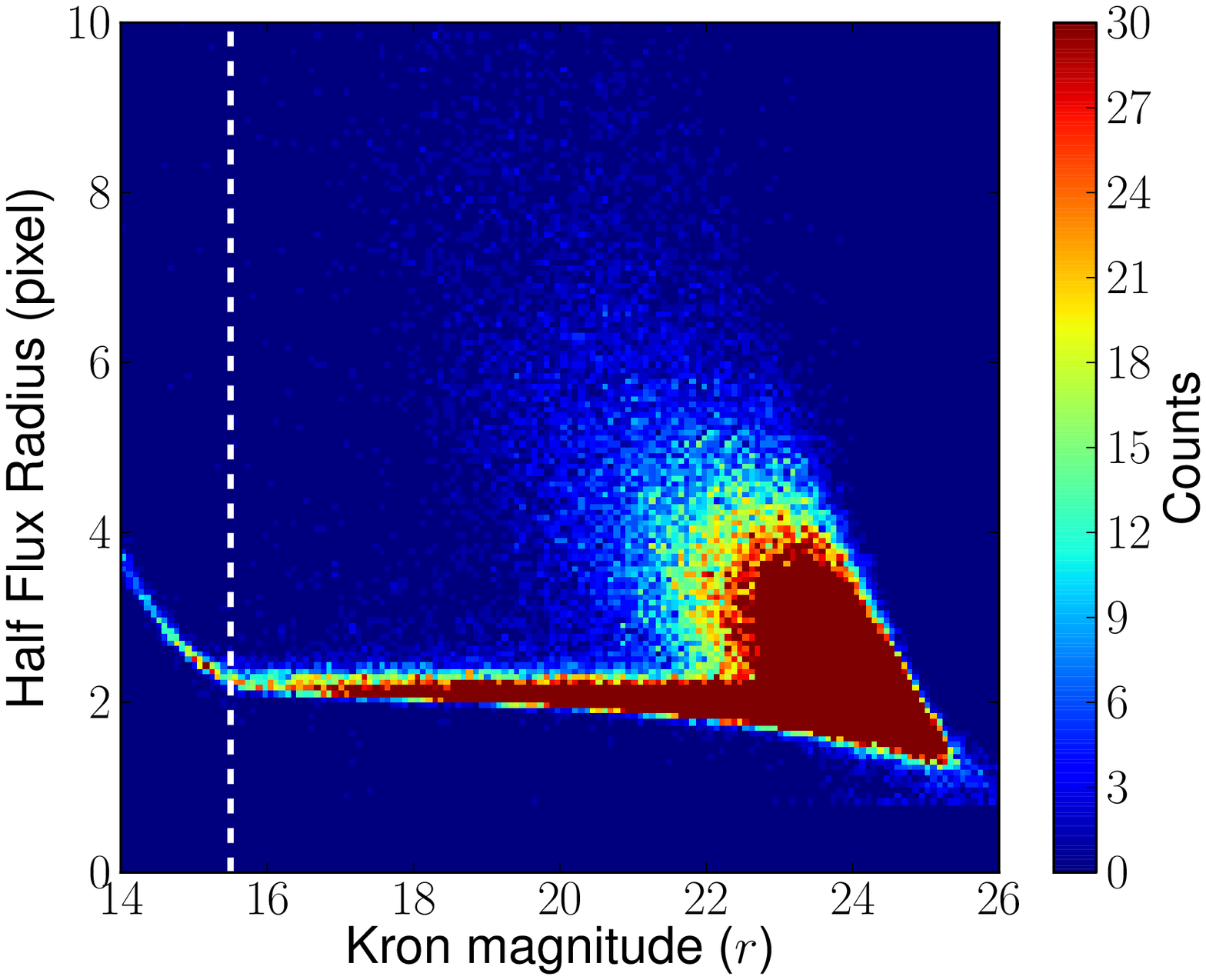}
\label{fig8b}}
\\
\subfloat[]{
\includegraphics[scale=0.4, bb= 10 170 540 595,clip]{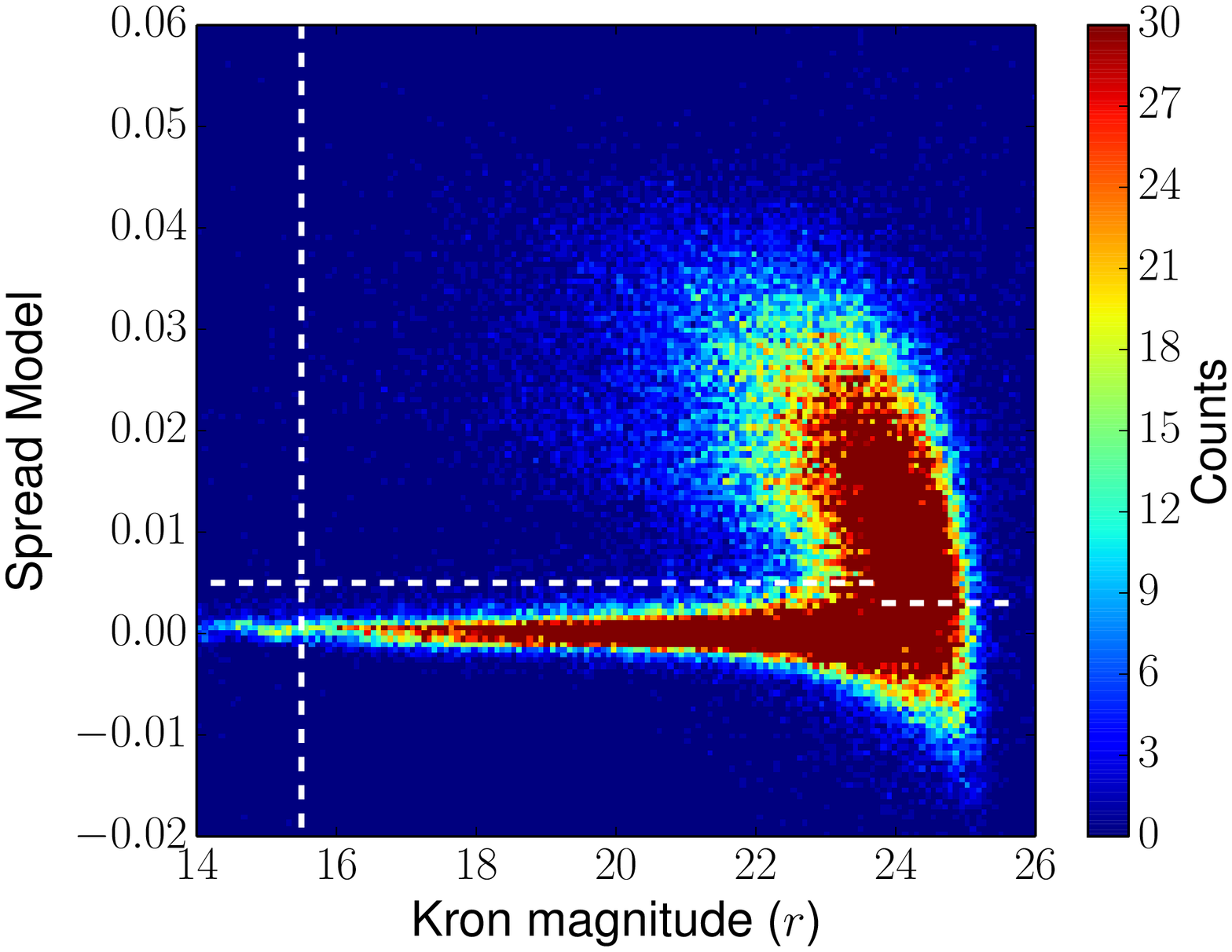}
\label{fig8c}}
\subfloat[]{
\includegraphics[scale=0.4, bb= 30 170 545 595,clip]{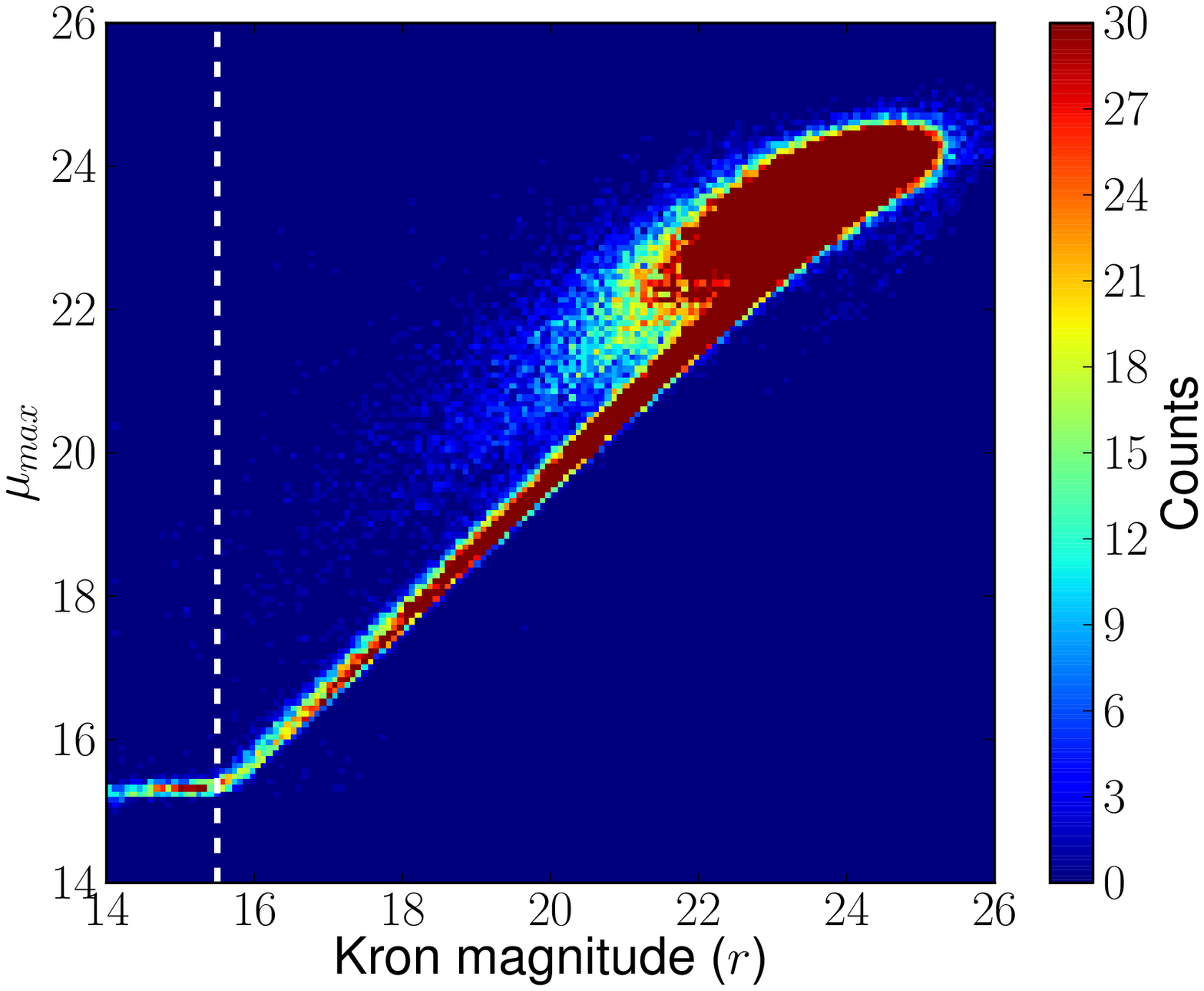}
\label{fig8d}}
\caption{Two-dimensional histogram of SExtractor stellarity index
  ({\it{panel a}}), half-light radius ({\it{panel b}}), spread model
  ({\it{panel c}}), and $\mathrm{\mu_{max}}$ ({\it{panel d}}) as a
  function of Kron magnitude for sources in $r$-band image of
  field 12. The vertical dashed line in all panels is the magnitude
  limit adopted for saturated objects. The horizontal lines in panel c) mark the limit used to separate stars and galaxies (see text). The points are colour coded
  according to their number counts as reported in the vertical
  colour bars.}
\end{figure*}

The \textit{stellarity index} results from a supervised neural network
that is trained to perform a star/galaxy classification. It assumes
values between 0 and 1. In theory, SExtractor considers objects with
{\tt{CLASS\_STAR}} equal to zero to be galaxies, and those with value
equal to one as stars.  As an example in Fig.~\ref{fig8a},
{\tt{CLASS\_STAR}} is plotted as a function of the Kron magnitude for
$r$-band sources in field 12. The sequence of unsaturated stars
($r>15.5$ mag) is clearly separated from galaxies by selecting a
{\tt{CLASS\_STAR}} value above 0.98 only down to $r$=20.5 mag. For
magnitudes fainter than this value, lowering the established limit to
separate stars and galaxies causes an increase of the contamination of
the star subsample from galaxies. So, we can only adopt this parameter
to classify bright sources ($r<$20.5 mag) as stars, when
{\tt{CLASS\_STAR}}$\ge$0.98.

By using ({\tt{FLUX\_RADIUS}}) as a measure of source
concentration, we can extend the classification to fainter
magnitudes. Figure~\ref{fig8b} shows that the locus of stars, defined
according to the relation between half-light radius and Kron
magnitude is recognizable down to $r$=22.8 mag. This limit is 0.7 mag
brighter than the completeness limit of the $r$-band catalogue (see
Sect.~\ref{sec:4.3}). For this reason we used the new SExtractor
classifier, {\tt{SPREAD\_MODEL}}, which takes into account the
difference between the model of the source and the model of the local
PSF \citep{desai12}, to obtain a reliable star/galaxy classification for the
faintest objects in our catalogue.  By construction,
{\tt{SPREAD\_MODEL}} is close to zero for point sources, positive for
extended sources (galaxies), and negative for detections smaller than
the PSF, such as cosmic rays. Figure~\ref{fig8c} shows the
distribution of the {\tt{SPREAD\_MODEL}} as a function of Kron
magnitude.  Stars and galaxies tend to arrange themselves in two
different places of the plot distinguishable up to $r$=23.8 that is a
limit 0.3 mag fainter than the completeness limit of the
catalogue. Based on this diagram, we classified as galaxies all
sources with SPREAD\_MODEL $>$ 0.005 and
$r \le $23.8, or SPREAD\_MODEL $>$ 0.003 and $r>$23.8.

Finally, in Fig.~\ref{fig8d} we plot $\mathrm{\mu_{max}}$ as a
function of the Kron magnitude. This plot is used in order to select
saturated stars (vertical dashed line in all panels of
Fig.~8).

Since the $r$ band is the deepest band of the survey, and the one conducted
in the best seeing conditions, it will be used for classification of
sources in the cross-correlated catalogue of the four bands. The same
criteria have also been used for the $i$ and $g$ band independently and
the results are consistent.

The star/galaxy separation of the $u$ band relies on the $r$ band,
since the nature of the $u$-band emission, which is very sensitive to
the presence of SF regions. For this band a careful cross-correlation
with the other catalogues is also required since the deblending
parameter cannot be suitably calibrated. We will further detail the
construction of the $u$-band catalogue in a dedicated article.

After completing the star/galaxy separation, a visual inspection of
those objects classified as galaxies but with
{\tt{CLASS\_STAR}}$\ge$0.9 (which are about 10 per field), is
performed.

To check the reliability of the star/galaxy classification in the
$i$-band images, we performed simulations by adding artificial stars
and galaxies, through a stepwise procedure. To preserve the
  overall source density, we split the artificial stars and galaxies
  into eight magnitude bins of width 0.5\,mag, over the magnitude range
  18$<i<$22, producing eight different simulated images for each real
  image. To define the PSF we take advantage of the PSF
modelled according to PSFEx. To simulate galaxies with dimensions
representative of real deep images, we defined the size in pixels of
the model in the output image, according to the distribution of the
measured half-flux radius as a function of the total magnitude
obtained for the real galaxies.
We attempted to recover artificial sources by running
SExtractor again with the same parameters for object detection and
classification as on the original images. We were able to correctly
separate 100\% of simulated stars and galaxies down to
$i$=20.0\,mag. We erroneously classify as galaxies 0.8\%, 1.5\%, 2.1\%
and 3.0\% of stars in the magnitude bins 20.0--20.5, 20.5--21.0,
21.0--21.5 and 21.5--22.0, respectively. All galaxies were correctly
recovered down to $i$=21.0, while in the last two magnitude bins just
0.7\% and 2.2\% were misclassified as stars.

To test if the method adopted to classify stars and galaxies could introduce a bias in the distribution of galaxy sizes we compared our observed distribution of real galaxy sizes, with those derived by using completely independent methods. In particular, we verified that our size distribution for galaxies down to M$^*$+6 was in agreement with those of D'Onofrio et al. (\citeyear{don14}) for WINGS clusters.

However, compact galaxies might be confused with stars due to their small radii and comparably high surface brightness. Previous studies on the sizes of dwarf and compact dwarf galaxies in Coma (e.g. Graham \& Guzman \citeyear{gra03}, Price et al. \citeyear{pri09}, Hoyos et al. \citeyear{hoy11}) and in other nearby clusters like Centaurus (Misgeld et al. \citeyear{mis09}), Antlia (Smith Castelli et al. \citeyear{ant12}), Virgo, Fornax and Perseus (Weinmann et al. \citeyear{wei11}) have shown that the effective radii of these galaxies can vary from few kiloparsec to hundreds of parsec. To investigate to what extent we are able to properly distinguish compact galaxies from stars we run {\it ad hoc} simulations. We added to the images artificial compact galaxies and used the same detection method as for the real catalogues. Simulated galaxies with a minimum effective radius of 0.7 kpc (i.e. $\sim$ 1 arcsec at z=0.048) are all recovered. Then, considering galaxies with a minimum effective radius of 0.6 kpc, we erroneously classify as stars only 2.2\% of compact galaxies. Finally, by simulating galaxies with effective radii down to 0.4 kpc we were able to classify as extended objects 60-70\% of galaxies. However, we would like to underline that 400 pc corresponds to $\sim$ 0.4 arcsec at the mean distance of the SSC (z=0.048), so, to assess the nature (star or galaxy) of these objects, we need either a measure of the redshift or to perform a fit to the surface brightness distribution.

\section{Photometry}
\label{sec:4}

In this section, we will focus on the photometric depth, the accuracy
of the derived photometry and the completeness magnitude.  All the
quantities shown in this section are obtained by excluding saturated
stars, sources affected by saturation spikes, stars in haloes of
bright stars, and those with less than half the total exposure of the
frames, which means that we included the sources with at least three
exposures in the {\it ugi} bands and with at least five exposures in the
{\it r} band. We remind the reader that the results for the $u$ band are
preliminary.

The results of Sections~\ref{sec:4.1} and \ref{sec:4.3} are based on the
$ugri$ catalogues including 11, 10, 14 and 23 fields,
respectively. For each band we cross-correlate the field catalogues
with STILTS selecting in the overlap regions the detection from the
image with the best seeing.  This criterion is also adopted for
producing the final $i$-band catalogue covering the whole ShaSS area.

We also report in Table~\ref{tab5} a summary of the exposure time,
median seeing, completeness limits and detection limits at 5$\sigma$
for each band.

\begin{table}
  \centering
\caption{ShaSS photometry.}
  \begin{tabular}{lllll}
    \hline \hline
Band & Exp. Time & completeness & detection [5$\sigma$] & seeing\,$^{a)}$  \\
     &   [s]     &      [mag]   & [mag]                 & [arcsec]\\
\hline
$u$ & 2955 & 23.8 & 24.4 & 0.9 \\

$g$ & 1400 & 23.8 & 24.6 & 0.7 \\

$r$ & 2664 & 23.5 & 24.1 & 0.6 \\

$i$ & 1000 & 22.0 & 23.3 & 0.7 \\
\hline
\hline
\end{tabular}
\begin{tabular}{l}
\end{tabular}
\begin{tabular}{l}
a) Median FWHM estimated from the already observed fields. \\
\end{tabular}
\label{tab5}
\end{table}

\subsection{Photometric depth}
\label{sec:4.1}

We estimated the photometric depth of each field by randomly placing
10000 3-$\arcsec$ apertures on the VST image of each
field. From the resulting standard deviation in the flux
  measurements obtained within these apertures, the corresponding
  20$\sigma$ and 5$\sigma$ magnitude limits were determined for each
  image. The mean values obtained in each field are 22.9, 23.1, 22.6,
21.7 in $ugri$ band at 20$\sigma$ and 24.4, 24.6, 24.1, 23.3 in $ugri$
band at 5$\sigma$. The variation of these detection limits across
  the fields is less than 0.05 in $ug$, and 0.1, 0.2 in $r$ and $i$
  band, respectively.  Merluzzi et al. (\citeyear{mer15}) reported
slightly different values since they measured the SNR within 3\,$\arcsec$ diameter apertures as a function of the
magnitude, according to equation (61) of SExtractor User's
Manual\footnote{\url{https://www.astromatic.net/pubsvn/software/sextractor/trunk/doc/sextractor.pdf}},
while in this paper we measured the detection limits directly on the
images.

\subsection{Photometric accuracy}
\label{sec:4.2}

To test the accuracy of our photometry we used the stars in the
3\,arcmin wide stripes of overlap between adjacent VST fields. The
magnitudes adopted for this comparison are those derived in an
8$\arcsec$ diameter aperture, which ensures that our estimates are not
affected by the seeing differences between fields. The average number
of stars in each strip are $\sim $650 in $u$ and 1500 in $gri$
bands.

We computed the magnitude difference $\Delta^{ij}_k=$m$^j_k$- m$^i_k$
as a function of mag$^{ij}_k=\frac{1}{2}\times$(m$^i_k$+m$^j_k$) for
all stars belonging to each pair $ij$ of adjacent frames. Here $i$ and
$j$ identify the different VST fields and $k$ refers to the stars, so
that m$^j_k$ is the magnitude of the $k$-th star in field $i$ and
$\Delta^{ij}_k$ is the difference in magnitude of the $k$-th star
between fields $j$ and $i$. 

To estimate the uncertainties on magnitudes, we collected the absolute
values of the differences $\left|\Delta^{ij}_k\right|$ and
mag$^{ij}_k$ of all stripes into two single vectors {\bf $\Delta_k$} and
{\bf mag$_k$}.
We then divided the sample into $N$ (=26) equally populated
(N$_{stars}\sim$1000-2000) bins of magnitude, and for each bin, we computed
the standard deviation of {\bf $\Delta_k$}, $\sigma_{bin}$, using a
3$\sigma$ rejection for the outliers (the subscript $bin$ refers to
the different bins in which the sample was divided). 

\begin{figure}
\centering
\includegraphics[scale=0.4]{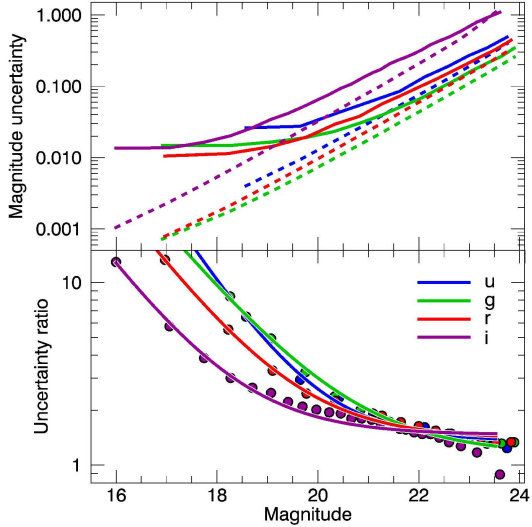}
\caption{{\it Top}: magnitude uncertainties derived from stars in the 3 arcmin
overlapping stripes between adjacent VST fields (continuous lines) and
uncertainties estimated by SExtractor (dashed) as functions of the
magnitude. The colours code the wavebands as indicated in the lower
panel. {\it Bottom}: ratio between the two uncertainty estimates. Continuous lines are the fit to the real data (filled circles) whose coefficients are reported in Tab~\ref{tab7}.}
\label{fig9}
\end{figure}

We adopt $\sigma_{bin}$ as the measure of the uncertainties of the
magnitude differences in each bin, and therefore $1/\sqrt{2}\times
\sigma_{bin}$ as an empirical measure of the uncertainties of the
magnitudes. These uncertainties are shown in the upper panel of Fig.~\ref{fig9}
as functions of mag$^{ij}_k$. The ratios of the empirical
uncertainties with those estimated by SExtractor are shown in the
lower panel of the same figure.

The ratios of the two uncertainties may be expressed as a function of
magnitude and band as
$$\Delta_{m} / \Delta_{SExtractor} = a_k \times exp(b_k \times m) +c_k  \ \ \ , $$
where $a_k, b_k, c_k $, (k=1-4) are coefficients depending on the
band which were derived by least-square fits to the curves in Fig.~\ref{fig9} (bottom).

The uncertainties given in the catalogue are those given by SExtractor
(i.e. the nominal error $\Delta_{SExtractor}$). To obtain a more
realistic error, the nominal error should be multiplied by $\Delta_{m}
/ \Delta_{SExtractor}$, with the zero-point uncertainty added then in
quadrature. In Tab.~\ref{tab7} we give the value of the multiplicative
factor as a function of the magnitude. Zero-point errors are reported in
Table~\ref{tab1}.

\begin{table}
\caption{Coefficients of the ratio between empirical and SExtractor uncertainties as a function of the magnitude - R=a$_1$$\times$exp[a$_2$ $\times$ (mag-20)]+a$_3$.}
  \label{tab_liter}
  \begin{center}
    \begin{tabular}{cccc}
\hline
Band & a$_1$ & a$_2$ & a$_3$\\
u & 1.26 & -0.985 & 1.353 \\
g & 18.55 & -0.762 & 1.144 \\
r & 9.39 & -0.839 & 1.397 \\
i & 3.67 & -0.858 & 1.467 \\
\hline
\hline
\end{tabular}
\label{tab7}
\end{center}
\end{table}

\subsection{Catalogue completeness}
\label{sec:4.3}

\begin{figure}
\centering
\includegraphics[scale=0.40, bb = 30 60 650 965,clip]{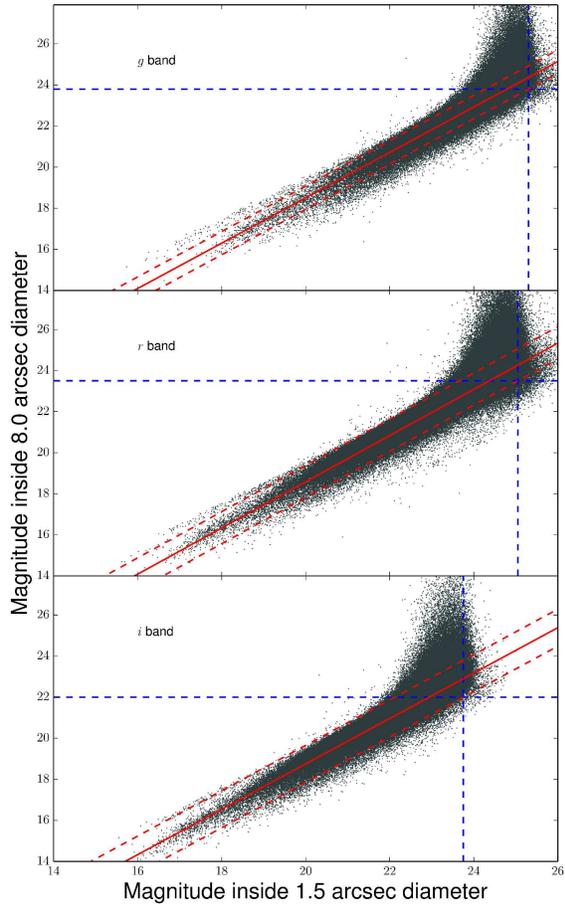}
\caption{Distribution of the SExtractor magnitude inside a
  8.0$^{\prime\prime}$ diameter as a function of the magnitude inside a
  detection aperture of 1.5$^{\prime\prime}$ diameter for $gri$ bands. The horizontal and vertical blue dashed lines indicate the detection and the completeness limits, respectively. The red continuous lines are the linear relation between the magnitude within 8.0$^{\prime\prime}$ aperture diameter and the magnitude within the detection aperture, minus/plus 1$\sigma$ (red dashed lines).}
\label{fig10}
\end{figure}

Following the method of \citet{gar99}, we estimated the completeness
magnitude limit as the magnitude at which we begin to lose galaxies
because they are fainter than the brightness threshold inside a
detection aperture of 1.5\,$\arcsec$ diameter\footnote{This
  aperture was adopted being suitable for all images and comparable to
  the larger seeing value.}, for the $g$, $r$ and $i$ bands. For the
$u$ band we give a first estimate of the completeness, based on the
distribution of the Kron magnitude, since the analysis of the
photometry is still ongoing. In all panels of Fig.~\ref{fig10} the
vertical blue dashed lines represent the detection limit, while the
red continuous lines are the linear empirical relation between the
magnitude within 8.0${\arcsec}$ diameter aperture and the magnitude
within the detection aperture. The relation between the two magnitudes
shows a scatter, depending essentially on the galaxy profiles. Taking
into account this scatter (see dashed red lines in Fig.~\ref{fig10}),
we fixed as a completeness magnitude limit (blue dashed horizontal
line) the intersection between the lower 1$\sigma$ limit of the
relation and the detection limit, which corresponds to 23.8,
  23.8, 23.5, 22.0 in $ugri$ bands, respectively. We checked the
  percentage of galaxies retrieved at the completeness limit defined
  above, by means of the recovery rate of artificial galaxies inserted
  in the images and retrieved with identical procedure as those used
  for real sources. We simulated 100 galaxies for each 0.5 mag
  bin over the magnitude range 18$<i<$23.5\,mag and effective radii
  ranging from 0.3 to 20\,arcsec. We verify that the catalogues turned
  out to be 93\% complete at the total magnitudes of 23.8, 23.8, 23.5,
  22.0 in $ugri$ band, respectively, and they reach the 89\% of
  completeness 0.5 mag fainter.

\begin{figure}
\centering
\includegraphics[scale=0.4, clip=true]{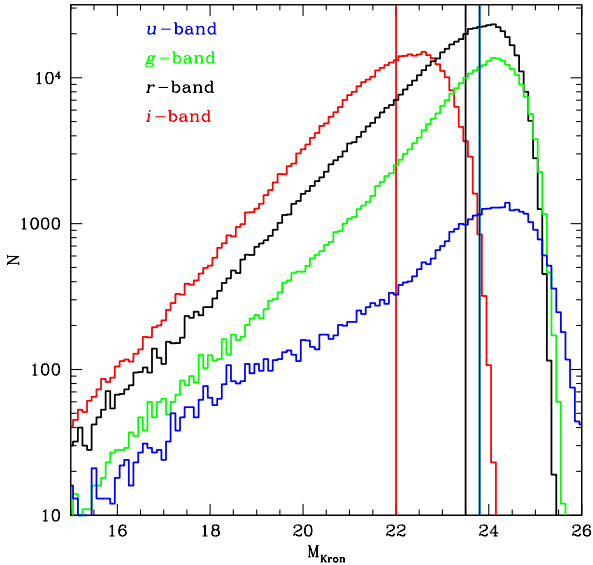}
\caption{Number counts of galaxies over the observed fields (11, 10, 14, 23, for $u$, $g$, $r$ and $i$, respectively) per 0.1 mag bin. Vertical lines mark the completeness magnitudes. $u$- and $g$-band catalogues have the same completeness magnitude indicated by the blue vertical line.}
\label{fig11}
\end{figure}

Fig.~\ref{fig11} shows the distribution of the Kron magnitude, where
vertical continuous lines indicate the limit to which the catalogues are
complete.

\section{Description of the data base}
\label{sec:5}

The $i$-band catalogue is published and it will be regularly
updated, through the use of the Virtual Observatory (VO) tools.

\begin{figure*}
\centering
\includegraphics[width=1.0\textwidth]{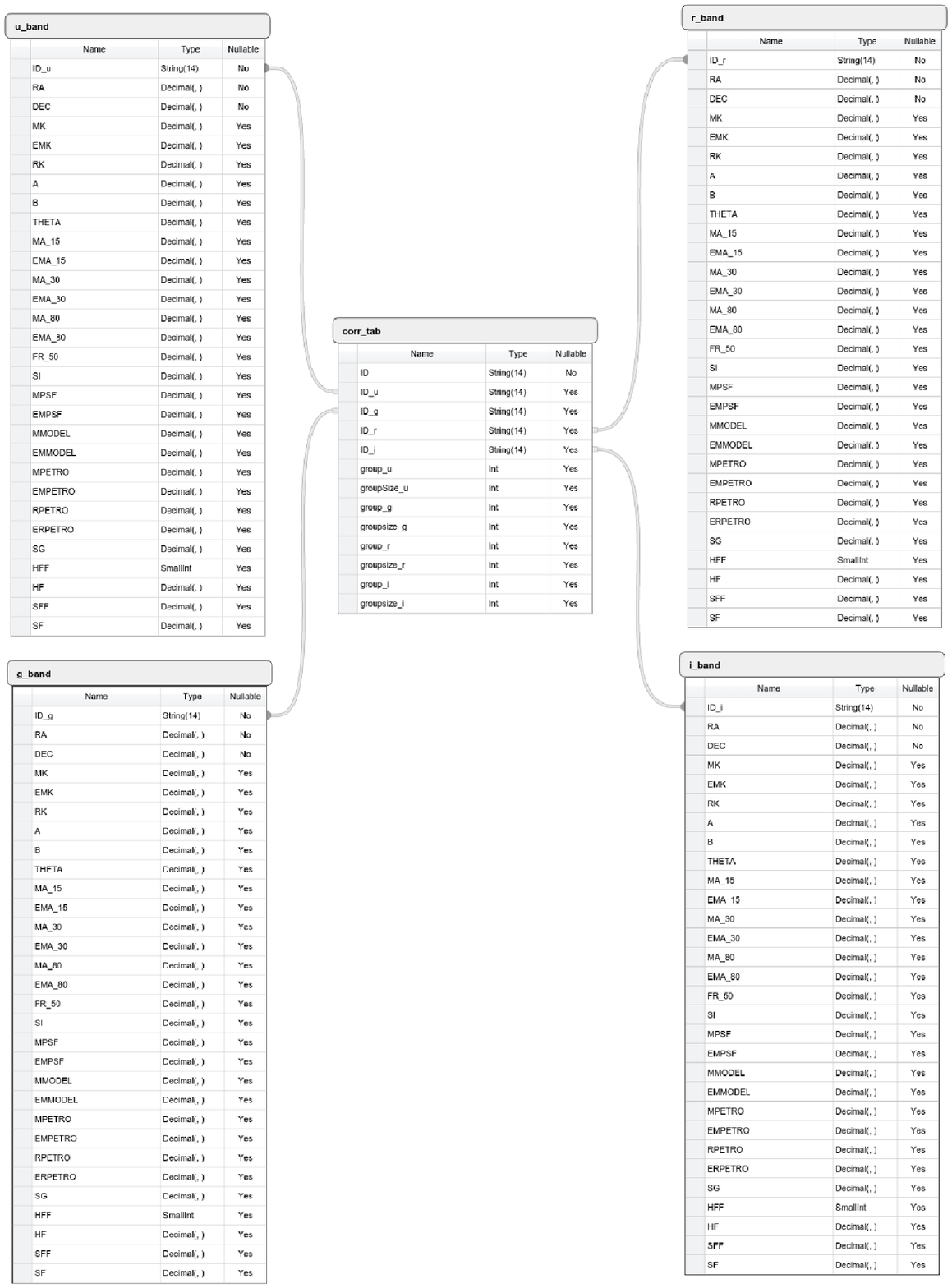}
\caption{The layout of the ER diagram for the optical ShaSS data base.}
\label{fig12}
\end{figure*}

The data model of the ShaSS data base is based on the standard
Entity-Relationship (ER) paradigm. Its architectural diagram is shown
in Fig.~\ref{fig12}. Its physical model (band-specific and
correlation tables) is instantiated in the logical DataBase Management
System (DBMS), based on the open source \textit{mySQL} data base and on
the \textit{SVOCat} service, a VO software tool
developed in the framework of the Spanish Virtual Observatory
(\url{http://svo2.cab.inta-csic.es/vocats/SVOCat-doc/)}\ .

The DBMS provides direct access to data through SQL-based queries and VO standard interoperability (i.e. direct table/image exchanging among VO tools, such as Topcat and Aladin). The service is publicly accessible via browser at the address \url{http://shass.na.astro.it}\ .

The Shass data base is also publicly available within the EURO-VO registry framework, under the INAF-DAME Astronomical Archive identification authority (\url{ivo://dame.astro.it/shass-i}).

Each object in the data base has a unique primary key
({\tt{SHASS\_ID}}), which univocally identifies sources. It is
composed by a string containing 14 characters, where five characters are
for \textit{ShaSS} and none digits \textit{NNNNNNNNN} represent the
internal object identification number.

The {\tt{SHASS\_ID}} field represents the primary identifier of an object within the corresponding band table. Whenever an object has a counterpart in different bands, the sequence of different {\tt{SHASS\_ID}}s for each reference band can be obtained by simple queries to the correlation table in the data base (for instance the corr\_tab table in Fig.~\ref{fig12}). In this table the fields {\tt{group}} and {\tt{group\_size}} are reported for each band. These indicate the possible presence of a group of objects matching with the single source within 1$\arcsec$ distance. To each group of sources matching with one specific object is assigned a unique integer, recorded in the {\tt{group}} field, and the size of each group is recorded in the {\tt{group\_size}} field. Sources which don't match any others (singles) have null values in both these fields.

In the data base we report the barycenter coordinates in degrees
({\tt{RAdeg}}, {\tt{DECdeg}}), the geometrical parameters of the
ellipse that describes the shape of the objects: semi-major and semi-minor axes,
and position angle ({\tt{A}}, {\tt{B}}, {\tt{THETA}}). As an indicator
of the area covered by the objects we include the half flux radius
({\tt{FR$_{50}$}}) which corresponds to the radius of the isophote
containing half of the total flux. We add the
Kron radius ({\tt{RK}}) and the Petrosian radius ({\tt{RPETRO}}),
which are the indicators of size of the Kron and the Petrosian
aperture respectively. Both of these radii are expressed in multiples
of the semi-major axis.

Among the extracted photometric quantities, we report in the public
catalogue three aperture magnitudes ({\tt{MA}}) (see Table~\ref{tab6}),
the Kron magnitude ({\tt{MK}}), the magnitude resulting from the PSF
fitting ({\tt{MPSF}}), the model magnitude obtained from the sum of
the spheroid and disk components of the fitting ({\tt{MMODEL}}), and
the Petrosian magnitude ({\tt{MPETRO}}). Magnitudes are not corrected
for galactic extinction and we give the relative uncertainties as
derived by SExtractor (but see also Sect.~\ref{sec:4.2}).

Finally we provide the stellarity index from SExtractor ({\tt{SI}})
and five flags: the star/galaxy flag ({\tt{SG}}), the halo fraction
flag ({\tt{HFF}}), the halo flag ({\tt{HF}}), the spike fraction flag
({\tt{SFF}}) and the spike flag ({\tt{SF}}).

The star/galaxy flag is fixed according to the star/galaxy separation
described in Sect.~\ref{sec:3.3}. Stars classified according to the
SExtractor stellarity index, half flux radius and the spread model
have {\tt{SG}}=1,2,3 respectively, while those classified through the
visual inspection have {\tt{SG}}=7.  Saturated stars are indicated by
{\tt{SG}}=9. Sometimes values of the star/galaxy flag {\tt{SG}}=4,5
could be present. They indicate stars aligned in a `secondary
sequence' visible in a few observed fields in the plots of the half
light radius and the $\mu_{max}$ versus the Kron magnitude,
respectively. The origin of this secondary sequence is unclear,
  but it seems a "random effect" since there is no correlation with
  observing night, airmass or spatial position of stars, and only
  constitutes a small number of stars.

As stated in Sect.~\ref{sec:3.3}, the $r$ band will be used for the
classification of sources in the cross-correlated catalogue, while in
the $g$ and $i$ single-band catalogues we report the
classification independently derived for each band, which is
mostly consistent with that of the $r$ band.

The halo and the spike fraction flag indicate the fraction of the
object area, defined as the circular area with radius equal to half
flux radius, affected by the presence of the halo or the spike of a
bright saturated star. Since, as pointed out in Sect.~\ref{sec:3.2}
the `intensity' of the halo is related to the magnitude of the parent
star, flag {\tt{HF}} is equal to 100$\times$ i$_{mag}$, where
i$_{mag}$ is the star magnitude from the USNO-B1.0 catalogue. Spikes
flag values are equal to 1,2 or 3, corresponding to three threshold
values of the surface brightness of 20, 20.5 and 21 mag arcsec$^{-2}$
of saturated regions (see Sect.~\ref{sec:3.2}).

A description of the units used for each of the quantities in
the data base is reported in Table~\ref{tab6}.

We plan to complement the present data base with the catalogues of the
$u$, $g$ and $r$ bands and the cross-correlated catalogue following
the completion of the observations.

\begin{table}
\centering
\small
\begin{tabular}{l l l}
\hline
\hline
Parameter & Units & Description\\
\hline
{\tt{ID}} & & ShaSS identification\\
{\tt{RAdeg}} & deg & Right ascension (J2000)\\
{\tt{DECdeg}} & deg & Declination (J2000)\\
{\tt{MK}} & mag & Kron magnitude\\
{\tt{EMK}} & mag & Error on Kron magnitude\\
{\tt{RK}} &  & Kron radius expressed in multiples of major axis\\
{\tt{A}} & deg & Major axis \\
{\tt{B}} &  deg & Minor axis \\
{\tt{THETA}} & deg & Position angle (CCW/x)  \\
{\tt{MA$_{15}$}} & mag & Aperture magnitude inside 1.5\,$\arcsec$ diameter\\
{\tt{EMA$_{15}$}} & mag & Error on aperture magnitude inside 1.5\,$\arcsec$ diameter\\
{\tt{MA$_{40}$}} & mag & Aperture magnitude inside 4.0\,$\arcsec$ diameter\\
{\tt{EMA$_{40}$}} & mag & Error on aperture magnitude inside 4.0\,$\arcsec$ diameter\\
{\tt{MA$_{80}$}} & mag & Aperture magnitude inside 8.0\,$\arcsec$ diameter\\
{\tt{EMA$_{80}$}} & mag & Error on aperture magnitude inside 8.0\,$\arcsec$ diameter\\
{\tt{FR$_{50}$}} & pixel & Radius of the isophote containing half of the total\\
                &       & flux\\
{\tt{SI}} & & SExtractor stellarity index\\
{\tt{MPSF}} & mag & Magnitude resulting from the PSF fitting\\
{\tt{EMPSF}} & mag & Error on magnitude resulting from the PSF fitting\\
{\tt{MMODEL}} & mag & Magnitude resulting from the model of the spheroid\\
              &     & and disc components \\
{\tt{EMMODEL}} & mag & Error on magnitude resulting from the model of the\\
               &     & spheroid and disc components\\
{\tt{MPETRO}} & mag & Petrosian magnitude\\
{\tt{EMPETRO}} & mag & Error on Petrosian magnitude\\
{\tt{RPETRO}} &  & Petrosian radius expressed in multiples of major\\
              &  & axis\\
{\tt{SG}} & & Star Galaxy separation\\
{\tt{HFF}} & & Halo fraction flag\\
{\tt{HF}} & & Halo flag value\\
{\tt{SFF}} & & Spike fraction flag\\
{\tt{SF}} & & Spike flag value\\
\hline
\end{tabular}
\caption{Parameters reported in the data base.}
\label{tab6}
\end{table}

\section{Summary}
\label{sec:6}

The ShaSS will map an area of $\sim$23 deg$^2$
($\sim$ 260 h$_{70}^{-2}$ Mpc$^2$ at z=0.048) of the SSC with the principal aim being to quantify the
influence of hierarchical mass assembly on galaxy evolution, and to
follow this evolution from filaments to cluster cores.

ShaSS will provide the first homogeneous multi-band imaging covering
the central region of the supercluster, including optical ($ugri$) and
NIR ($K$) imaging acquired with VST and VISTA, which allows accurate
multiband photometry to be obtained for the galaxy population down to
m*+6 at the supercluster redshift. In particular, the $r$-band images
are collected with a median seeing equal to 0.6\,arcsec, corresponding
to 0.56 kpc h$^{-1}_{70}$ at z=0.048 and thus enabling the internal properties of
supercluster galaxies to be studied and distinguish the impact of the
environment on their evolution.

In this article, we described the methodology for producing
photometric catalogues. The optical survey is ongoing and the analysis
presented in this work is performed on 11, 10, 14 and VST
fields out of 23 in $ugr$ bands, nevertheless in all bands the considered
subsamples are representative of the final whole sample. The
$i$-band imaging is instead complete.

The catalogues are produced using the software SExtractor
\citep{bertin96} in conjuction with PSFEx (\citealp{bertin11}), and a
careful analysis of the software outputs. 

We were able to obtain a robust separation between stars and galaxies
up to the completeness limit of the optical data, through a
progressive approach using: i) the \textit{stellarity index}
({\tt{CLASS\_STAR}}); ii) the half-light radius ({\tt{FLUX\_RADIUS}})
; iii) the new SExtractor classifier {\tt{SPREAD\_MODEL}}; iv) the
peak of the surface brightness above background
(\texttt{$\mathrm{\mu_{max}}$}); v) a final visual inspection for
objects classified as galaxies but with
{\tt{CLASS\_STAR}}$>$0.90. 

The ShaSS catalogues reach average 5$\sigma$ limiting magnitudes
inside a 3$\arcsec$ aperture of $ugri$=[24.4,24.6,24.1,23.3] and a
completeness limit of $ugri$=[23.8,23.8,23.5,22.0], which corresponds
to $\sim$ m*$_r$+8.5 at the supercluster redshift. These values correspond
to the survey expectations.

The $i$-band catalogue is released to the community through the use of
the Virtual Observatory tools. It includes 734 319 sources down to
$i$=22.0 over the whole area. The catalogue is 93\% complete
at this magnitude limit and 34\% of the sources are galaxies.
  The service is publicly accessible via
browser at the address \url{http://shass.na.astro.it}\ . The Shass
data base is also publicly available within the EURO-VO registry
framework, under the INAF-DAME Astronomical Archive identification
authority (\url{ivo://dame.astro.it/shass-i})\ .

\section*{Acknowledgements}

The authors thank the referee, R. Smith, for his constructive comments
and suggestions. This work is based on data collected with the ESO - VLT Survey
Telescope with OmegaCAM (ESO Programmes 088.A-4008, 089.A-0095,
090.A-0094, 091.A-0050) using Italian INAF Guaranteed Time
Observations. The data base has made use of SVOCat, a VO publishing
tool developed in the framework of the Spanish Virtual Observatory
project supported by the Spanish MINECO through grant AYA $2011-14052$
and the CoSADIE FP7 project (Call $INFRA-2012-3.3$ Research
Infrastructures, project $312559$). SVOCat is maintained by the Data
Archive Unit of the Centro de Astrobiología (CSIC -INTA). The research
leading to these results has received funding from the European
Community’s Seventh Framework Programme (FP7/2007-13) under grant
agreement number 312430 (OPTICON; PI: P. Merluzzi) and PRIN-INAF 2011:
{\it Galaxy evolution with the VLT Surveys Telescope (VST)} (PI
A. Grado). CPH was funded by CONICYT Anillo project ACT-1122. PM
thanks M. Petr-Gotzens for her support in the VST observations.  AM
and MB acknowledges financial support from PRIN-INAF 2014: {\it Glittering
Kaleidoscopes in the sky, the multifaceted nature and role of galaxy
clusters} (PI M. Nonino). PM and GB acknowledge financial support from
PRIN-INAF 2014: {\it Galaxy Evolution from Cluster Cores to Filaments} (PI
B.M. Poggianti).


\label{lastpage}

\end{document}